\documentclass[12pt,a4paper]{article}

\usepackage[T1]{fontenc} 
\usepackage{verbatim}
\usepackage[utf8]{inputenc}
\usepackage[american]{babel}
\usepackage{graphicx}
\usepackage{epsfig}
\usepackage{adjustbox}
\usepackage{booktabs}
\usepackage{multirow}
\usepackage{dcolumn}
\usepackage{amsmath}
\usepackage{amsfonts}
\usepackage{mathtools}
\usepackage{amssymb}
\usepackage{subcaption} 
\usepackage{caption}
\usepackage{epstopdf}
\usepackage{bm}
\usepackage{mathrsfs}
\usepackage{version}
\usepackage{enumerate}
\usepackage{braket}
\usepackage{enumitem}
\usepackage{transparent}
\usepackage{pifont}
\usepackage{colortbl}
\usepackage{rotating}
\usepackage{adjustbox}
\usepackage{cancel}
\usepackage{tabularx}
\usepackage{soul}
\usepackage[table]{xcolor}

\usepackage{hyperref}
\hypersetup{colorlinks=true,linkcolor=royalblue,citecolor=magenta,pdfencoding=auto}

\definecolor{navyblue}{rgb}{0.0, 0.0, 0.5}
\definecolor{royalblue}{rgb}{0.25, 0.41, 0.88}
\definecolor{cadmiumgreen}{rgb}{0.0, 0.42, 0.24}
\definecolor{blue-violet}{rgb}{0.54, 0.17, 0.89}
\definecolor{darkviolet}{rgb}{0.58, 0.0, 0.83}
\definecolor{orange(colorwheel)}{rgb}{1.0, 0.5, 0.0}

\definecolor{magenta(process)}{rgb}{1.0, 0.0, 0.56}

\definecolor{darkspringgreen}{rgb}{0.09, 0.45, 0.27}

\definecolor{royalblue(web)}{rgb}{0.25, 0.41, 0.88}

\definecolor{cadmiumorange}{rgb}{0.93, 0.53, 0.18}

\definecolor{heliotrope}{rgb}{0.87, 0.45, 1.0}

\newcommand\Coe{{\cal A}}

\makeatletter
\renewcommand*{\@textcolor}[3]{%
\protect\leavevmode
\begingroup
\color#1{#2}#3%
\endgroup
}
\makeatother

\renewcommand\({\left(}
\renewcommand\){\right)}
\renewcommand\[{\left[}

\makeatletter
\let\save@mathaccent\mathaccent
\newcommand*\if@single[3]{%
\setbox0\hbox{${\mathaccent"0362{#1}}^H$}%
\setbox2\hbox{${\mathaccent"0362{\kern0pt#1}}^H$}%
\ifdim\ht0=\ht2 #3\else #2\fi
}
\newcommand*\rel@kern[1]{\kern#1\dimexpr\macc@kerna}
\newcommand*\widebar[1]{\@ifnextchar^{{\wide@bar{#1}{0}}}{\wide@bar{#1}{1}}}
\newcommand*\wide@bar[2]{\if@single{#1}{\wide@bar@{#1}{#2}{1}}{\wide@bar@{#1}{#2}{2}}}
\newcommand*\wide@bar@[3]{%
\begingroup
\def\mathaccent##1##2{%
\let\mathaccent\save@mathaccent
\if#32 \let\macc@nucleus\first@char \fi
\setbox\z@\hbox{$\macc@style{\macc@nucleus}_{}$}%
\setbox\tw@\hbox{$\macc@style{\macc@nucleus}{}_{}$}%
\dimen@\wd\tw@
\advance\dimen@-\wd\z@
\divide\dimen@ 3
\@tempdima\wd\tw@
\advance\@tempdima-\scriptspace
\divide\@tempdima 10
\advance\dimen@-\@tempdima
\ifdim\dimen@>\z@ \dimen@0pt\fi
\rel@kern{0.6}\kern-\dimen@
\if#31
\overline{\rel@kern{-0.6}\kern\dimen@\macc@nucleus\rel@kern{0.4}\kern\dimen@}%
\advance\dimen@0.4\dimexpr\macc@kerna
\let\final@kern#2%
\ifdim\dimen@<\z@ \let\final@kern1\fi
\if\final@kern1 \kern-\dimen@\fi
\else
\overline{\rel@kern{-0.6}\kern\dimen@#1}%
\fi
}%
\macc@depth\@ne
\let\math@bgroup\@empty \let\math@egroup\macc@set@skewchar
\mathsurround\z@ \frozen@everymath{\mathgroup\macc@group\relax}%
\macc@set@skewchar\relax
\let\mathaccentV\macc@nested@a
\if#31
\macc@nested@a\relax111{#1}%
\else
\def\gobble@till@marker##1\endmarker{}%
\futurelet\first@char\gobble@till@marker#1\endmarker
\ifcat\noexpand\first@char A\else
\def\first@char{}%
\fi
\macc@nested@a\relax111{\first@char}%
\fi
\endgroup
}
\makeatother

\newcommand{\mal}[1]{\mathcal #1}

\newcommand{\expect}[1]{\left\langle #1 \right\rangle}

\newcommand\ee{\end{equation}}
\newcommand\be{\begin{equation}}
\newcommand\eea{\end{eqnarray}}
\newcommand\bea{\begin{eqnarray}}
\newcommand{\bsp}{\begin{split}}
\newcommand{\esp}{\end{split}}
\newcommand{\bit}{\begin{itemize}[leftmargin=*]}
\newcommand{\eit}{\end{itemize}}
\newcommand{\ben}{\begin{enumerate}[leftmargin=*]}
\newcommand{\een}{\end{enumerate}}

\renewcommand{\arraystretch}{1.2}

\newcommand\eq[1]{Eq.~\eqref{eq:#1}}

\newcommand{\dif}{\mathrm{d}}

\renewcommand{\vec}{\bm} 

\newcommand{\cs}{c_\mathrm{s}}

\def\O{\mathcal{O}}

\def\k{\vec{k}}

\def\d{\partial}
\def\ep{\epsilon}

\def\tr{{{^{(3)}}\!R}}

\def\dn{\delta N}
\def\de{\delta E}
\def\Rf{{{^{(4)}}\!R}}
\def\VV{V}
\def\AA{A}
\def\ggg{\gamma\gamma\gamma}
\def\ggz{\gamma\gamma\zeta}

\def\mpl{M_{\rm Pl}}

\begin{document}
\begin{center}
\LARGE{\textbf{Simplifying the EFT of Inflation: \\ Generalized Disformal Transformations \\ and Redundant Couplings }} \\
\end{center}
\vspace{0.3cm}

\begin{center}

\large{Lorenzo Bordin,$^{\rm a,b}$ Giovanni Cabass,$^{\rm c}$ Paolo Creminelli$^{\rm d}$ and Filippo Vernizzi$^{\rm e}$}
\\[0.7cm]

\small{
\textit{$^{\rm a}$ SISSA, via Bonomea 265, 34136, Trieste, Italy}}
\\[0.1cm]
\small{
\textit{$^{\rm b}$ INFN, National Institute for Nuclear Physics, Via Valerio 2, 34127 Trieste, Italy}}
\\[0.1cm]
\small{ \textit{$^{\rm c}$ Physics Department and INFN, Universit\`a di Roma ``La Sapienza'', \\ P.le\ Aldo Moro 2, 00185, Rome, Italy}}
\\[0.1cm]
\small{
\textit{$^{\rm d}$ Abdus Salam International Centre for Theoretical Physics, \\ Strada Costiera 11, 34151, Trieste, Italy}}
\\[0.1cm]
\small{
\textit{$^{\rm e}$ Institut de physique th\'eorique, Universit\'e Paris Saclay, CEA, CNRS, 91191 Gif-sur-Yvette, France}}

\vspace{.1cm}

\end{center}

\vspace{.8cm}

\hrule
\vspace{0.3cm}
\noindent\small{\textbf{Abstract}\\ 
\noindent We study generalized disformal transformations, including derivatives of the metric, in the context of the Effective Field Theory of Inflation. All these transformations do not change the late-time cosmological observables but change the coefficients of the operators in the action: some couplings are effectively redundant. At leading order in derivatives and up to cubic order in perturbations, one has $6$ free functions that can be used to set to zero $6$ of the $17$ operators at this order. This is used to show that the tensor three-point function cannot be modified at leading order in derivatives, while the scalar-tensor-tensor correlator can only be modified by changing the scalar dynamics. At higher order in derivatives there are transformations that do not affect the Einstein-Hilbert action: one can find 6 additional transformations that can be used to simplify the inflaton action, at least when the dynamics is dominated by the lowest derivative terms. We also identify the leading higher-derivative corrections to the tensor power spectrum and bispectrum.

\vspace{0.3cm}
\noindent
\hrule
\vspace{.8cm}

\section{Introduction}
\label{sec:intro}

\noindent In scalar-tensor theories one is used to conformal transformations of the metric and the possibility to describe physics in different frames. When dealing with backgrounds in which the scalar field is time-dependent, like in the case of inflation, one can consider disformal transformations \cite{Bekenstein:1992pj} of the form
\be
g_{\mu \nu} \to {\cal C}(\phi,X) g_{\mu \nu} + {\cal D}(\phi,X) \partial_\mu \phi \partial_\nu \phi \;, \qquad X \equiv -(\partial\phi)^2 \;.
\ee
In a gauge in which the inflaton perturbations are set to zero ($\phi = \phi_0(t)$), the so-called unitary gauge, these transformations are written as
\be
\label{confdisf}
g_{\mu \nu} \to C(t,N) g_{\mu \nu} + D(t,N) n_\mu n_\nu \;,
\ee
where $N \equiv (-g^{00})^{-1/2}$ is the lapse and $n^\mu$ is a unit vector perpendicular to the surfaces of constant inflaton, $n_\mu \equiv {- \partial_\mu} \phi/\sqrt{X}$, and $C(t,N)$ and $D(t,N)$ can be easily related to ${\cal C}(\phi,X)$ and ${\cal D}(\phi,X)$. 
However this is not the end of the story. More general transformations are possible if one considers objects with derivatives on the metric: 
\be
\label{general_transf}
g_{\mu \nu} \to C(t,N,V,K, \tr, \dots) g_{\mu \nu} + D(t,N,V, K,\tr, \dots) n_\mu n_\nu + E(t,N,V, K,\tr, \dots) K_{\mu\nu} + \dots \;,
\ee
where $K_{\mu\nu}$ is the extrinsic curvature of the surfaces of constant inflaton, $K = g^{\mu\nu}K_{\mu\nu}$, and $\tr$ their scalar curvature. 
The purpose of this paper is to study these general transformations in the context of the Effective Field Theory of Inflation (EFTI) \cite{Creminelli:2006xe,Cheung:2007st} and to understand to which extent they can be used to simplify the original action. 
This generalizes the results of \cite{Creminelli:2014wna}, where disformal transformations with time dependent coefficients $C(t)$ and $D(t)$ were used to remove, without loss of generality, the time-dependence of the Planck mass and a non-trivial speed of tensor modes in the EFTI. The Planck mass and the tensor speed are couplings that can be changed without affecting observables: in QFT these are called {\em redundant couplings}.

When calculating S-matrix elements, field redefinitions cannot change the final result. 
In cosmology one is interested in correlation functions 
and, contrarily to S-matrix elements, 
these are not invariant under field redefinitions.
However, {\em late-time} correlation functions---the ones that are relevant for observations---are left invariant by the transformations discussed above. 
Indeed, at late time all derivatives of metric perturbations decay to zero and the lapse gets its background value, $N \to 1$. 
We are left with a transformation of the form $g_{\mu \nu} \to C(t) g_{\mu \nu} + D(t) n_\mu n_\nu $. This redefines the scale factor and the cosmic time of the background FRW solution, but scalar and tensor perturbations are not changed.
Therefore, {\em the general transformation eq.~\eqref{general_transf} modifies the form of the action, without changing late-time correlators}. 
The identification of the minimal set of non-redundant operators in the context of inflation was carried out in \cite{Weinberg:2008hq}, albeit with some different assumptions. 

The effect of general disformal transformations on the EFTI operators have been also studied in the context of dark energy and modified gravity in \cite{Gleyzes:2015pma,DAmico:2016ntq,Langlois:2017mxy}. 
In this case one is interested also in the way matter couples with the metric and this coupling is  modified by the redefinition of the metric. On the other hand, in single field models of inflation  the coupling with matter does not enter in the  inflationary predictions and therefore we will not consider it in the following.

Since we are talking about an infinite set of possible field redefinitions, an organization principle is needed. 
In the EFTI (and in all other EFTs!) one organizes operators in terms of order of perturbations (tadpoles, quadratic terms, cubic terms, etc.) and in a derivative expansion. 
The same can be done for the transformations.
Transformations involving derivatives of the metric, such as $K_{\mu\nu}$ and $\tr$, will increase the number of derivatives in the action. 
For instance, starting from the Einstein-Hilbert term, they will generate operators with three or more derivatives.
Therefore, let us start with transformations without derivatives on the metric. 
An additional simplification comes from expanding the transformations in powers of perturbations. 
Since we dropped all terms with more derivatives, this boils down to an expansion in powers of $\delta N$. 
If one is interested in correlation functions up to cubic order---like we are in this paper---one can truncate the transformations to quadratic order in perturbations. 
Indeed transformations which are cubic in perturbations will only modify the action with terms that are at least quartic, since the field redefinition multiplies the equations of motion, which vanish at zeroth order in perturbations. 
Therefore, one is left with
\be
\label{f1234}
g_{\mu \nu} \to \left(f_1(t) + f_3(t)\delta N + f_5(t)\delta N^2 \right) g_{\mu \nu} + \left(f_2(t) + f_4(t)\delta N+ f_6(t)\delta N^2 \right) n_\mu n_\nu \;.
\ee

This set of transformations will change the coefficients of the operators in the EFTI. 
In particular one can consider all the operators with at most two derivatives on the metric (and up to cubic order in perturbations) and study which simplifications are allowed by the six free functions $f_i(t)$: we are going to do that in Section \ref{sec:operators} and Appendix \ref{app:transformations}. 
One can use the free functions to set to zero the coefficients of the operators in the EFTI. 
The choice of which operator should be set to zero is clearly arbitrary. However, since only a few operators enter in calculations involving tensor modes, a natural choice is to try and simplify as much as possible the tensor couplings. 
The functions $f_1$ and $f_2$ can be used to have a time-independent Planck mass and to set to unity the speed of tensor modes \cite{Creminelli:2014wna}. This procedure also fixes the correlator $\langle\gamma\gamma\gamma\rangle$ at leading order in derivatives. 
In Section \ref{conti_gamma_gamma_zeta} we are going to see that the functions $f_3$ and $f_4$ can similarly be used to simplify the coupling $\gamma\gamma\zeta$ in such a way that the correlator $\langle\gamma\gamma\zeta\rangle$ is only modified by changes in the scalar sector. In particular we are going to verify that different actions, related by eq.~\eqref{f1234}, give the same result for $\langle\gamma\gamma\zeta\rangle$. 
In Appendix~\ref{app:transformations} we explicitly give the effect of a general transformation, eq.~\eqref{f1234}, on operators up to two derivatives on the metric. 
The transformations $f_5$ and $f_6$ cannot be used to standardize any coupling involving tensors, so that their use to set to zero some operators remains, to some extent, arbitrary.

While the field redefinitions of Section~\ref{sec:operators} do not change the number of derivatives in the operators, in Section~\ref{sec:higherder} we consider transformations that add one or more derivatives. In particular, in Section~\ref{diff-like_trans} we study a subset of the field redefinitions \eqref{general_transf} that act on the metric as diffeomorphisms and leave the Einstein-Hilbert action invariant. 
Provided that higher derivatives can be treated perturbatively, we can use six of these transformations to further reduce the number of independent operators.
In Sections \ref{subs:tensor_higher} and Appendix \ref{zeta_gamma_gamma_bispectrum} we study higher-derivative corrections to tensors. 
We show that there is a single operator with three derivatives that modifies $\expect{\gamma\gamma\gamma}$ and a single operator that modifies the tensor power spectrum at 4-derivative order. We calculate the contribution of this operator to $\braket{\gamma\gamma\zeta}$: it is not slow-roll suppressed and therefore potentially relevant.
Finally, in Section \ref{sec:decoup_limit} we discuss additional field redefinitions that one can perform in the decoupling limit. These are not necessarily constrained by the nonlinear realization of Lorentz invariance and we show that they cannot be generally recovered by field transformations in unitary gauge.
Conclusions are drawn in Section~\ref{conclusions}.

\section{Operators up to two derivatives}
\label{sec:operators}

\noindent We want to consider the action of the EFTI up to second order in derivatives and cubic in perturbations \cite{Creminelli:2006xe,Cheung:2007st}. 
In order to do so, we introduce the ADM decomposition of the metric \cite{Arnowitt:1962hi}, i.e.
\be
\label{ADM}
\dif s^2 = -N^2 \dif t^2 + h_{ij} (\dif x^i + N^i \dif t) (\dif x^j + N^j \dif t) \;,
\ee
in terms of the lapse $N$, the shift $N^i$ and the spatial metric $h_{ij}$. For later use we decompose the latter as \cite{Maldacena:2002vr}
\be
h_{ij} = a^2(t) e^{2 \zeta} ( e^\gamma )_{ij} \;, \qquad \gamma_{ii}=0 \;,
\ee
and we define the Hubble rate, $H \equiv \dot a/a$. 
The unitary gauge can be fixed by choosing the time coordinate to coincide with constant inflaton hypersurfaces and by imposing $\partial_i \gamma_{ij} =0$ on these slices.
In this gauge, $\zeta$ and $\gamma_{ij}$ respectively represent the scalar and tensor propagating degrees of freedom. 

The unitary gauge EFTI action reads
\be
\label{general_action}
S = S_0 + \int\dif^4x\,\sqrt{-g} \left( {\cal L}^{(2)}+ {\cal L}^{(3)} + \dots \right) \;,
\ee
where $S_0$ is the minimal canonical action \cite{Cheung:2007st}
\be
\label{eq:EFT_action-sfsr}
S_0\equiv \frac{\mpl^2}{2}\int \dif^4x\,\sqrt{-g}\bigg[\Rf - \frac{2\dot H}{N^2} - 2 (3H^2 + \dot{H})\bigg] \;,
\ee
and
\be\label{EFTI_operators}
{\cal L}^{(2)} \equiv M_{\rm Pl}^2 \sum_{I=0}^{8} a_I (t) {\cal O}_I^{(2)} \;, \qquad {\cal L}^{(3)} \equiv M_{\rm Pl}^2 \sum_{I=1}^{8} b_I(t) {\cal O}_I^{(3)} \;,
\ee
are, respectively, linear combinations of $9$ quadratic operators, ${\cal O}^{(2)}_I$ ($I=0,\dots,8$), and $8$ cubic operators ${\cal O}^{(3)}_I$ ($I=1,\dots,8$). 
The list of quadratic operators is given in Table \ref{table:quadratic}, while the cubic ones in Table \ref{table:cubic}. 
The operators in these tables are constructed by combining the perturbation of the lapse, $\delta N \equiv N - 1$, of the extrinsic curvature, $\delta K_{\mu\nu} \equiv K_{\mu\nu} - H h_{\mu\nu}$, and of its trace $\delta K \equiv K - 3 H$. 
The 3-dimensional Ricci scalar curvature $\tr$ is already a perturbed quantity, because we are assuming a flat FRW background. Moreover, the ``acceleration’’ 
vector $\AA^\mu$ is given by $\AA^\mu \equiv n^\nu \nabla_\nu n^\mu$: it 
is projected on the surfaces of constant inflaton, i.e. $\AA^\mu n_\mu=0$, and can also be written as $\AA_\mu = N^{-1}\, h^{\nu}_{\,\,\mu}\partial_\nu N$. 
With $V$ we denote the covariant derivative of the lapse projected along $n^\mu$, 
\be
\label{Vdef}
V \equiv n^\mu \nabla_\mu N = \frac1{N} \left( \delta \dot N - N^i \partial_i N \right) \;, 
\ee
which is a 3-dimensional scalar. Indeed, using the unitary gauge relation $n^\mu = - N g^{\mu 0}$, $V$ is proportional to the upper time derivative of the lapse, i.e.~$V = - N \partial^0 N$. 
Operators like $\delta N V$ and $\delta N^2 V$ can be written in terms of $\delta N$ and $\delta K$ after integrations by parts, and we can always get rid of $R^{00}$ using the Gauss-Codazzi relation of eq.~\eqref{GC2} below. 
In Table \ref{table:quadratic} and Table \ref{table:cubic} we also indicate the number of derivatives of each operator and whether an operator modifies 
a given coupling: only a few operators modify couplings that include gravitons.\footnote{Notice that scalar operators such as $\delta N$, $\delta K$ and $V$ cannot contain $\gamma_{ij}$ at linear order in perturbations.}
For the time being, we do not assume any hierarchy among these operators (while in Sec.~\ref{sec:higherder} we will).

\renewcommand{\arraystretch}{1.2}
\begin{table}
\begin{center}
\begin{adjustbox}{max width=\textwidth}
\begin{tabular}{ | c | l | c | c | c | c | c | c | c | c | }
\hline 
Coeff.	& ${\cal O}^{(2)}$					& \#$\partial_\mu$	& $\gamma\gamma$	& $\gamma\gamma\gamma$	& $\gamma\gamma\zeta$	& $\gamma\zeta\zeta$	& $\zeta\zeta\zeta$	& $\to 0$		\\ [0.0cm] 
\hline 
\hline 
$a_0$	& $\tr$							&2			& $\checkmark$	& $\checkmark$			& $\checkmark$		& $\checkmark$		& $\checkmark$	& $f_{1,2}$	\\ [0.0cm] 
\hline 
$a_1$	& $\delta K_{\mu\nu}\delta K^{\mu\nu}$	& 2			& $\checkmark$	& $\checkmark$			& $\checkmark$		& $\checkmark$		& $\checkmark$	& $f_{1,2}$	\\ [0.0cm] 
\hline 
$a_2$	& $\tr\,\delta N$						& 2			&				&						& $\checkmark$		& $\checkmark$		& $\checkmark$	& $f_{3,4}$	\\ [0.0cm] 
\hline 
$a_3$	& $\AA_\mu \AA^\mu$				& 2			&				&						&					& $\checkmark$		& $\checkmark$	&			\\ [0.0cm] 
\hline 
$a_4$	& $H^2 \delta N^2$					& 0			&				&						&					&					& $\checkmark$	&			\\ [0.0cm] 
\hline 
$a_5$	& $H \delta N \delta K$				& 1			&				&						&					&					& $\checkmark$	&			\\ [0.0cm] 
\hline 
$a_6$	& $\delta K^2$						& 2			&				&						&					&					& $\checkmark$	&			\\ [0.0cm] 
\hline 
$a_7$	& $\VV^2$						& 2			&				&						&					&					& $\checkmark$	&			\\ [0.0cm] 
\hline 
$a_8$	& $\VV \delta K$					& 2			&				&						&					&					& $\checkmark$	&			\\ [0.0cm] 
\hline
\end{tabular}
\end{adjustbox}
\end{center}
\caption{\footnotesize{Quadratic operators up to second order in derivatives, 
together with the list of the couplings they affect. 
The last column shows which transformation can be used to set to zero the corresponding operator.}}
\label{table:quadratic}
\end{table}

\renewcommand{\arraystretch}{1.2}
\begin{table}
\begin{center}
\begin{adjustbox}{max width=\textwidth}
\begin{tabular}{ | c | l | c | c | c | c | c | c | c | c | }
\hline
Coeff.	& ${\cal O}^{(3)}$						& \#$\partial_\mu$	& $\gamma\gamma\gamma$	& $\gamma\gamma\zeta$	& $\gamma\zeta\zeta$	& $\zeta\zeta\zeta$	& $\to 0$		\\ [0.0cm] 
\hline
\hline 
$b_1$	& $\delta N\delta K_{\mu\nu}\delta K^{\mu\nu}$	& 2			&						& $\checkmark$		& $\checkmark$		& $\checkmark$	& $f_{3,4}$	\\ [0.0cm] 
\hline
$b_2$	& $\tr\,\delta N^2$						& 2			&						&					&					& $\checkmark$	& $f_{5,6}$	\\ [0.0cm] 
\hline
$b_3$	& $\delta N\AA_\mu \AA^\mu$				& 2			&						&					&					& $\checkmark$	& $f_{5}$		\\ [0.0cm] 
\hline
$b_4$	& $H^2 \delta N^3$						& 0			&						&					&					& $\checkmark$	&			\\ [0.0cm] 
\hline
$b_5$	& $H \delta N^2 \delta K$					& 1			&						&					&					& $\checkmark$	& $f_{5,6}$	\\ [0.0cm] 
\hline 
$b_6$	& $\delta N \delta K^2$					& 2			&						&					&					& $\checkmark$	&			\\ [0.0cm] 
\hline
$b_7$	& $\delta N\VV^2$						& 2			&						&					&					& $\checkmark$	&			\\ [0.0cm] 
\hline
$b_8$	& $\delta N \VV \delta K$					& 2			&						&					&					& $\checkmark$	& $f_{5}$		\\ [0.0cm]
\hline
\end{tabular}
\end{adjustbox}
\end{center}
\caption{\footnotesize{Cubic operators up to second order in derivatives, 
together with the list of the couplings they affect. The last column shows which transformation can be used to set to zero the corresponding operator. Two of the operators among $b_2$, $b_3$, $b_5$ and $b_8$ can be se to zero using the transformations $f_5$ and $f_6$.}}
\label{table:cubic}
\end{table}

Let us now study how one can use the transformations eq.~\eqref{f1234} to simplify the action. In \cite{Creminelli:2014wna} the transformations $f_1$ and $f_2$ were used to make the quadratic action for gravitons canonical. This boils down to eliminate the first two operators in Table \ref{table:quadratic} in such a way that the spatial and time kinetic term of the graviton only arise from the standard Einstein-Hilbert action with time-independent $\mpl$. Since the transformations $f_1$ and $f_2$ do not contain perturbations, they cannot be done perturbatively. They also modify the background FRW and the definition of cosmic time. The details are spelled out in \cite{Creminelli:2014wna} and in Appendix \ref{app:transformations}. 
The bottom line is that there is no loss of generality in setting 
to zero the first two operators in Table \ref{table:quadratic}. 
Notice that, since only these two operators modify the coupling $\gamma\gamma\gamma$, one concludes that the correlator $\langle\gamma\gamma\gamma\rangle$ cannot be modified at leading order in derivatives. We come back to corrections at higher order in derivatives in Section \ref{conti_gamma_gamma_gamma}.

Consider now the transformations of order ${\cal O}(\delta N)$, i.e.~$f_3$ and $f_4$. At leading order in perturbations the field redefinitions multiply the variation of the action with respect to the metric, i.e.~the equations of motion. 
In particular, the variation of the Einstein-Hilbert action under the transformation $g_{\mu \nu} \to g_{\mu \nu} + \delta g_{\mu \nu} $ gives
\be
\label{EH_trans}
\delta S_{\rm EH} = \frac{\mpl^2}{2}\int \dif^4x\,\sqrt{-g} \,G_{\mu \nu} \delta g^{\mu \nu} \;,
\ee
which for the transformations $f_3$ and $f_4$ becomes
\be
\label{EHf3andf4}
\delta S_{\rm EH} = \frac{\mpl^2}{2}\int\dif^4x\,\sqrt{-g} \left({f_3}\Rf\,\delta N - f_4 \delta N G_{\mu\nu} n^\mu n^\nu \right) \;.
\ee
Using the geometric 
Gauss-Codazzi relations \cite{Wald:1984rg} 
\begin{subequations}
\begin{align}
&\Rf = \tr - K^2 + K_{\mu\nu}K^{\mu\nu} + 2\nabla_\mu(K n^\mu - \AA^\mu)\;, \label{GC1}\\
&G_{\mu\nu}n^\mu n^\nu = \frac{1}{2}\left(\tr + K^2 - K_{\mu\nu}K^{\mu\nu}\right)\;, \label{GC2}
\end{align}
\end{subequations}
one can write the variation of the action 
in the EFTI form. 
We postpone all details to the next section: here it suffices to notice that one can use $f_3$ and $f_4$ 
to set the operators $\tr\,\delta N$ and 
$\delta N \delta K_{\mu\nu} \delta K^{\mu\nu}$ to zero. This choice 
can be convenient
since these are the only (remaining) operators that modify the coupling $\gamma\gamma\zeta$. In the following section we are going to verify explicitly the invariance of the correlator $\langle\gamma\gamma\zeta\rangle$ in doing the transformations $f_3$ and $f_4$.

The logic is the same for the two functions $f_5$ and $f_6$. Since the operators they generate are proportional to $\delta N^2$ and there is no scalar that one can build at linear order that contains $\gamma$, $f_5$ and $f_6$ do not affect anything that has to do with tensor modes at this order. Therefore, the choice of which operator to set to zero with $f_5$ and $f_6$ is, to some extent, arbitrary.
In Appendix~\ref{app:transformations} we explicitly calculate the variation of the operators under the transformations $f_i$ and we find which ones can be set to zero, see last column of Tables \ref{table:quadratic} and \ref{table:cubic}.
In particular, $f_5$ can set to zero one of the following 2-derivative operators: $\tr\,\delta N^2$, $\delta N \VV \delta K$, $H\delta N^2 \delta K$ or $\delta N \AA_\mu \AA^\mu$, while $f_6$ only $\tr\,\delta N^2$ or $H\delta N^2 \delta K$. 

Although in this paper we focus on terms that are up to cubic in perturbations, one can easily see what happens at higher order in $\delta N$. At each new order one gets a table similar to Table \ref{table:cubic} with more powers of $\delta N$. These are $8$ new operators at each order. At the same time one has $2$ new possible field redefinitions of the same form of $f_5$ and $f_6$ but with more powers of $\delta N$. One can use these two free functions to set to zero two of the new operators. 
In conclusion, one remains with $6$ non-redundant operators at each order in perturbations.

\subsection{Simplifying \texorpdfstring{$\expect{\gamma\gamma\zeta}$}{<\textbackslash gamma\textbackslash gamma\textbackslash zeta>}}
\label{conti_gamma_gamma_zeta}

\noindent As an explicit application and check, in this section we will show how to exploit the field redefinitions \eqref{f1234} to set to zero the operators involving two gravitons and a scalar. 
As shown in the tables, these operators are
\begin{equation}
\label{eq:ggz_operators}
\tr\;, \qquad \delta K_{\mu\nu}\delta K^{\mu\nu}\;, \qquad \tr\,\dn\;, \qquad \dn \delta K_{\mu\nu}\delta K^{\mu\nu}\;.
\end{equation}
All these operators already appear in the Einstein-Hilbert action. 
As explained in Section~\ref{sec:operators}, transformations $f_1$ and $f_2$ can be used to remove $\tr$ and $\delta K_{\mu\nu}\delta K^{\mu\nu}$ \cite{Creminelli:2014wna}.
We will verify that the redefinitions
\begin{subequations}
\label{eq:f3_and_f4_transformations}
\begin{align}
&g_{\mu\nu} \rightarrow g_{\mu\nu} + f_3\dn g_{\mu\nu}\;, \label{eq:f3_and_f4_transformations-1} \\
&g_{\mu\nu} \rightarrow g_{\mu\nu} + f_4\dn n_\mu n_\nu\;, \label{eq:f3_and_f4_transformations-2}
\end{align}
\end{subequations}
can be used to set to zero, respectively, $\tr\,\dn$ and $\dn \delta K_{\mu\nu}\delta K^{\mu\nu}$. 
In particular, we will show that the action $S_0$, eq.~\eqref{eq:EFT_action-sfsr}, changes under the transformations \eqref{eq:f3_and_f4_transformations},
but the late-time correlation functions do not.\footnote{We assume that the time dependence of the parameters $f_3$ and $f_4$ is mild enough. More precisely, 
$f_3$ and $f_4$ must not grow faster than $\eta^{-2}$ for $\eta\to 0$.} 

It is important to stress that, although the coupling $\gamma\gamma\zeta$ can be brought back to the standard Einstein-Hilbert form, the correlator $\expect{\gamma\gamma\zeta}$ is also sensitive to the solution of the scalar constraints. As such the correlator is modified by the quadratic scalar operators and it is not completely fixed at the 2-derivative level, contrarily to what happens for $\expect{\gamma\gamma\gamma}$.

Before proceeding, it is convenient to remind how the ADM components of the metric \eqref{ADM} change under the metric transformation of eq.~\eqref{confdisf} \cite{Gleyzes:2014qga}:
\be
\label{transfNNh}
h_{ij} \to C (t,N) h_{ij} \;, \qquad N^2 \to \left[C (t,N) - D (t,N)\right]N^2 \;, \qquad N^i \to N^i \;.
\ee

\subsubsection{Transformation \texorpdfstring{$f_4$}{f\_4}}
\label{sec:f_4_transformation}

\noindent We start by considering the disformal transformation $f_4$, eq.~\eqref{eq:f3_and_f4_transformations-2}, which is the simplest to treat. 
We can work at linear order in the metric transformation, because higher orders carry two or more powers of $\delta N$ and hence do not contribute to $\expect{\gamma\gamma\zeta}$. 
To keep calculations simple we assume $|f_{3,4}|\ll 1$ and constant in time.

The logic will be the following. By eq.~\eqref{transfNNh}, $f_4$ only affects the lapse: the action for $\zeta$ and $\gamma_{ij}$ has to be invariant under this transformation once the Hamiltonian and momentum constraints are solved. 
This is not obvious, because the intermediate action that explicitly contains the lapse changes. 
However, also the relation between the lapse and $\zeta$ given by the solution of the constraints changes accordingly.
We will check that once the lapse is replaced in terms of $\zeta$, the action for $\zeta$ and $\gamma_{ij}$ remains unchanged by the transformation. 

To do this, let us study the variation of the action $S_0$, eq.~\eqref{eq:EFT_action-sfsr}, under the transformation $f_4$. 
For the Einstein-Hilbert part of the action, one can use the Gauss-Codazzi relation \eqref{GC2} in eq.~\eqref{EHf3andf4}. 
Adding the variation of the scalar part, one finds 
\begin{equation}
\label{eq:f_4-S_variation}
\delta S_0 = - \frac{\mpl^2}{2}\int\dif^4x\,\sqrt{-g}\,f_4\delta N \bigg[ \frac{1}{2} \left(\tr + K^2 - K_{ij}K^{ij} \right) + \dot{H} \left({\frac{1}{N^2}} - 1 \right) - 3H^2 \bigg]\;. 
\end{equation}
The coefficient $f_4$ can be used, for example, to set to zero the operator $\tr \, \delta N$.
To verify that the action in terms of $\zeta$ and $\gamma$ does not change, we need to solve the lapse in terms of $\zeta$ only at linear order. 
This is because its second-order part does not contribute to the cubic action, as it multiplies the background equations of motion \cite{Maldacena:2002vr}. 
Thus, we can focus on the quadratic action. 
To do so, it is convenient to define $E_{ij} \equiv N K_{ij}$, whose explicit components are
\be
E_{ij} \equiv \frac{1}{2} \left( \dot h_{ij} - D_i N_j - D_i N_j \right) \;,
\ee
where $D_i$ denotes the covariant derivative with respect to the spatial metric $h_{ij}$.
With this notation and using that $\sqrt{-g} = N\sqrt{h} = a^3 e^{3 \zeta} N $, we can expand $S_0$ and $\delta S_0$ above at quadratic order. 
They read, respectively, 
\be
\label{eq:f_3-quadratic_action_zeta-A}
\begin{split}
S_0^{(2)} = \frac{\mpl^2}{2} \int\dif t\,\dif^3 x \,a^3\bigg[{-2} (3 H^2 + \dot H)\dn^2 + 4H\dn\delta E + \delta E ^i_{\,\,j}\delta E^j_{\,\,i} - \delta E ^2 + \tr\,\dn + 3 \tr\,\zeta \bigg]\;,
\end{split}
\ee
where $\delta E$ denotes the trace of $\delta E_{ij} \equiv E_{ij} - H h_{ij}$, and
\begin{equation}
\label{eq:f_4-full_quadratic_variation}
\delta S_0^{(2)} = - \frac{\mpl^2}{2}\int\dif t\,\dif^3 x \, a^3 \,f_4\bigg[\frac{1}{2}\tr\,\dn - 2 (3 H^2 + \dot H) \dn^2 + 2H\dn\de + 18 H^2 \delta N \zeta \bigg]\;.
\end{equation}
Varying $S^{(2)}_0 + \delta S^{(2)}_0$ with respect to the shift yields the momentum constraint, 
\be
D_i \left[ (E^i_{\ j} - E \delta^i_{\ j})N^{-1} - f_4 H N \delta N \delta^i_{\ j} \right] = 0 \;.
\ee
Solving this equation for the lapse, gives
\begin{equation}
\label{eq:f_4-shift_constraint_solved}
\dn = \frac{\dot\zeta}{H}\bigg(1+\frac{f_4}{2}\bigg)\;.
\end{equation}
At this point, it is straightforward to verify that plugging the above expression for $\dn$ in the original action $S_0$, the term proportional to $f_4$ which is generated exactly cancels the action variation \eqref{eq:f_4-S_variation}. 
We have also checked that the expression of the shift in terms of $\zeta$, which can be obtained from the Hamiltonian constraint, is not modified by the disformal transformation.

\subsubsection{Transformation \texorpdfstring{$f_3$}{f\_3} }
\label{sec:f_3_transformation}

\noindent The conformal transformation $f_3$, eq.~\eqref{eq:f3_and_f4_transformations-1}, is more complicated than the previous one, because not only does it changes the solution for $\dn$ but it also redefines $\zeta$. 
Indeed, working again at linear order in the metric transformation, from eq.~\eqref{transfNNh} we find the following transformations for $\zeta$ and $\delta N$: 
\be
\label{eq:dof_change}
\zeta\to\zeta + f_3\frac{\dn}{2} \;, \qquad \dn\to\dn\bigg(1+ \frac{f_3}{2}\bigg) \;, 
\ee
while the scalar component of the shift, defined as $\psi \equiv \partial^{-2}\partial_i N^i$, remains unchanged. 
The solutions of the Hamiltonian and momentum constraints change accordingly. 
Using these transformations in the usual solutions for $\dn$ and $\psi$ derived from action $S_0$ and assuming for simplicity a constant $f_3$, these are given respectively by
\begin{align}
&\dn = \frac{\dot\zeta}{H} - \frac{f_3}{2} \bigg(\frac{\dot{\zeta}}{H} - \frac{1}{H}\frac{\dif}{\dif t}\frac{\dot{\zeta}}{H}\bigg)\;, \label{deltaNzetaf3} \\
&\psi = {-\frac{{\zeta}}{a^2H}} + \ep\d^{-2}\dot{\zeta} - \frac{f_3}{2}\bigg[\frac{1}{a^2H}\frac{\dot{\zeta}}{H} - \ep\d^{-2}\bigg(\frac{\dif}{\dif t}\frac{\dot{{\zeta}}}{H}\bigg)\bigg] \;, \qquad \ep \equiv -\frac{\dot{H}}{H^2} \;. \label{eq:f3_dof_change-3}
\end{align}

Let us also derive these two relations by solving the constraints of the new action. For the Einstein-Hilbert part of the action, one can use again the Gauss-Codazzi relation \eqref{GC1} in eq.~\eqref{EHf3andf4}. 
Integrating by parts, using the definitions of $A_i = N^{-1} \partial_i N$ and $V$ (eq.~\eqref{Vdef}), and adding the variation of the scalar part, one finds 
\begin{equation}
\label{eq:f_3_variation}
\begin{split}
&\delta S_0 = - \frac{\mpl^2}{2}\int\dif^4x\,\sqrt{-g}\,f_3 \bigg\{ \delta N \bigg[ K^2 - K_{ij}K^{ij} - \tr + 2 \dot{H} \left(\frac{1}{N^2} +2 \right) + 12 H^2 \bigg] \\
&\hphantom{\delta S_0 = \frac{\mpl^2}{2}\int\dif^4x\,\sqrt{-g}\,f_3 \bigg\{ } + K V - A_i A^i N^{-1} \bigg\} \;.
\end{split}
\end{equation}
Here the lapse appears with a time derivative in $V$, which makes $\dn$ dynamical. 
This can be also seen by varying the action $S_0 + \delta S_0$ with respect to the shift. 
One obtains
\be
D_i\left\{ ({E^i_{\,\,j} - E \delta^i_{\,\,j}}){N^{-1}} -f_3 \left[ {-\dn} ({E^i_{\,\,j} - E \delta^i_{\,\,j}}){N^{-1}} + V \delta^i_{\,\,j} \right]\right\} - f_3 E A_j = 0\;,
\ee
which is a dynamical equation for $\delta N$ and not a constraint. However, since $V$ comes only at first order in $f_3$ it can be treated perturbatively. 
Indeed, solving this equation perturbatively in $f_3$ one recovers eq.~\eqref{deltaNzetaf3}. 
Moreover, the Hamiltonian constraint equation derived from this action gives the solution of eq.~\eqref{eq:f3_dof_change-3} for the shift.

The transformation $f_3$ changes the quadratic action for scalar perturbations (but not the one of tensors). This implies that the correlation functions $\expect{\zeta\zeta}$ and $\expect{\gamma\gamma\zeta}$ change when evaluated inside the horizon. Only at late times, the correlation functions will not depend on $f_3$ as we are now going to show. 
Let us first look at the quadratic action for scalar perturbations to verify that this is the case for the two-point function of $\zeta$. The second-order expansion of the action $S_0$ is given by eq.~\eqref{eq:f_3-quadratic_action_zeta-A}. Expanding the action \eqref{eq:f_3_variation} at second order yields, after some integrations by parts, 
\be
\label{eq:f_3-quadratic_action_zeta-B}
\begin{split}
\delta S_0^{(2)} = \mpl^2\int\dif^4x\,f_3\,a^3 \bigg[&{\frac{\tr\,\dn}{2}} - \frac{5}{2}H^2(3-\ep)\dn^2 - 2H\dn\delta E + 3 HN^i\d_i\dn \\
&- \delta\dot{N}\delta E + \frac{(\d_i\dn)^2}{a^2} - 9 (3 H^2 + \dot H)\dn \zeta - 3H\delta\dot{N} \zeta\bigg]\;.
\end{split}
\ee
Expressing the action above as function of the curvature perturbation $\zeta$ using eqs.~\eqref{deltaNzetaf3} and \eqref{eq:f3_dof_change-3}, the second-order expansion of $S_0 + \delta S_0$ gives 
\be
\label{eq:f_3-quadratic_action_zeta-C}
S^{(2)}_\zeta = {M}_\mathrm{Pl}^2 \left(1 - \frac{3f_3}{2} +\O(f_3^2)\right) \int\dif^4x\,a^3\epsilon\bigg[\dot\zeta^2 - \cs^2\frac{(\d_i \zeta)^2}{a^2}\bigg]\;,
\ee
with 
\be
\label{Mplcs}
\cs = 1 + \frac{f_3}{2} + \O(f_3^2) \;.
\ee 
Therefore, both the normalization and the speed of propagation of $\zeta$ are affected by the transformation $f_3$. 
This is reflected in a change of the wavefunction for $\zeta$, which becomes
\be
\label{eq:zeta_wave}
\zeta(\eta,k) = \frac{-i H}{2\sqrt{\ep}\, {M}_\mathrm{Pl} (1-3f_3/4) } \frac{1}{(\cs k)^{3/2}}(1+ i \cs k\eta) e^{-i\cs k\eta}\;.
\ee
However, the late-time two-point function of $\zeta$ does not change. 
Indeed, this is proportional to $(1+3f_3/2) \cs^{-3} = 1$ at leading order in $f_3$.

Let us now move to the computation of the cubic action $\gamma\gamma\zeta$. 
After many integrations by parts that show that the action is slow-roll suppressed \cite{Maldacena:2002vr}, we obtain
\begin{equation}
\label{eq:f_3-cubic_action_ggz}
\begin{split}
S^{(3)}_{\gamma\gamma\zeta} = \frac{\mpl^2}{4} \int\dif^4x\,a^3\ep\bigg\{&2\zeta\bigg(\dot\gamma_{ij}^2 + \frac{(\d_k\gamma_{ij})^2}{a^2}\bigg) - \dot\gamma_{ij}\d_k\gamma_{ij}\frac{\d_k}{\d^2}\dot\zeta \\
&- {f_3}\bigg[{-\frac{1}{4}}\frac{\dot\zeta}{H}\bigg(\dot\gamma_{ij}^2 + \frac{(\d_k\gamma_{ij})^2}{a^2}\bigg) + \frac{1}{2}\dot\gamma_{ij}\d_k\gamma_{ij}\frac{\d_k}{\d^2}\bigg(\frac{\dif}{\dif t}\frac{\dot\zeta}{H}\bigg)\bigg]\bigg\}\;.
\end{split}
\end{equation}
We can thus compute the $\expect{\gamma\gamma\zeta}$ three-point function. 
To do this, we need to use the wavefunctions of eq.~\eqref{eq:zeta_wave} in the in-in calculation. 
The final result is independent of $f_3$ up to $\O(f_3^2)$ corrections, 
thus confirming that late-time correlation functions are insensitive to the transformation of eq.~\eqref{eq:f3_and_f4_transformations-1}.

\section{Transformations of higher order in derivatives}
\label{sec:higherder}

So far, we have considered only field redefinitions without derivatives. These do not change the number of derivatives of the operators. 
In this section we consider more general transformations \eqref{general_transf}, involving one or two derivatives on the metric.

\subsection{Diff-like transformations}
\label{diff-like_trans}

In general, field redefinitions which include derivatives will generate, starting from the Einstein-Hilbert actions operators with 3 or more derivatives absent from Tables \ref{table:quadratic} and \ref{table:cubic}.
However there is a particular set of higher-derivatives transformations which do not change the Einstein-Hilbert action but only the inflaton one. 
Indeed, consider the transformation 
\be
\label{diff-like_field_redef}
g_{\mu \nu} \to g_{\mu \nu} + \nabla_{\mu} \xi_{\nu} + \nabla_{\nu} \xi_{\mu}\;,
\ee
where $\xi_\mu$ is a vector field starting linear in perturbations. This is analogous to a linearized diffeomorphism generated by $\xi_\mu$ (notice, however, that we are \emph{not} reintroducing the Stueckelberg $\pi$). The Einstein-Hilbert action does not change under this transformation because it is invariant under 4-dimensional diffeomorphisms. 
Since the EFTI action \eqref{general_action} is invariant under spatial diffeomorphisms, 
in the following we will consider only transformations along $ n_\mu$, which are associated to time-diffeomorphisms. Note that eq.~\eqref{diff-like_field_redef} is a particular case of the general transformation \eqref{general_transf}, as it can be checked by replacing in the above equation the general expression for $\xi_\mu$ in unitary gauge, i.e.,
\be
\label{eff}
\xi_\mu = F(t,N, V, K, \ldots) \, n_\mu \;.
\ee

We can thus focus on how the rest of the action transforms. We assume that the inflaton dynamics is dominated by $\delta N^2$ and $\delta N^3$ with the other operators (that we list in Table \ref{table:3}) being suppressed by negative powers of some energy scale $\Lambda$. (Our assumptions do not apply to cases in which the quadratic action of the inflaton is dominated by higher-derivative operators, such as, for instance, Ghost Inflation \cite{ArkaniHamed:2003uz} and Galileons \cite{Nicolis:2008in}.)
In this case the size of the operators is parametrically governed by a derivative expansion in $\partial_\mu/\Lambda$ and the coefficients in front of the higher-derivative operators are suppressed by positive powers of $H/\Lambda$. 
More specifically, if operators with no derivatives are of the order of the slow-roll parameter $\epsilon = -\dot H /H^2$, $a_4,b_4 \sim {\cal O}(\epsilon) $, those with one derivative are suppressed by a single power of $H/\Lambda$, $a_5,b_5 \sim {\cal O}(\epsilon H/\Lambda)$, while those with two derivatives are suppressed by $ (H/\Lambda)^2$. 
Since the above transformation generates at least one more derivative in the action,
the variation of the operators with two derivatives is suppressed by at least $(H/\Lambda)^3$ and we neglect it here. 

We first focus on transforming the operators with no derivatives. In this case, it is straightforward to compute the variation of the action \eqref{general_action} under eq.~\eqref{diff-like_field_redef}. 
In particular, we use that the transformation of the lapse is given by
\be
\delta_\xi N = -N n_\mu n_\nu \nabla^\mu \xi^\nu \;.
\ee 
Assuming that the operator coefficients $a_I$ and $b_I$ are time independent and neglecting slow-roll corrections, the linear variation of the action \eqref{general_action} reads, up to third order in perturbations,
\be\label{matter_var_diff-like-2}
\begin{split}
&\delta_\xi S = -M_{\rm Pl}^2\int\dif^4x \sqrt{-g} \, \bigg\{ 2 \dot H \left( \frac{3H}{N} +\frac{V}{N^3} -\frac{ K }{N^2} \right) + 2 a_4 H^2 \Big[ 3H (1+\delta N )\delta N \\
&\hphantom{\delta_\xi S = -M_{\rm Pl}^2\int\dif^4x \sqrt{-g} \, \bigg\{ } + (1+3 \delta N) V + \delta N \delta K \Big] + 3 b_4 H^2 (3 H \delta N + 2 V ) \delta N \bigg\}\, F\;,
\end{split}
\ee
with $F$ defined by eq.~\eqref{eff}. 

Restricting it to be at most first order in derivatives, we have
\be
F = \frac1H \left[ g_1(t) \delta N + g_2(t) \delta N^2 + g_3 (t)\frac{V}{H} + g_4 (t)\delta N\, \frac{V}{H}+ g_5 (t) \frac{\delta K}{H} + g_6 (t) \delta N\, \frac{\delta K}{H} \right]\;,
\ee
where, typically, $g_1$ and $g_2$ are suppressed by $H/\Lambda$ while $g_{3,4,5,6}$ carry a $(H/\Lambda)^2 $ suppression. From eq.~\eqref{matter_var_diff-like-2} one sees that the transformations $g_1$ and $g_2$ generate one-derivative operators suppressed by $H/\Lambda$. Thus, they
can be used 
to set to zero the operators $a_5 \delta N \delta K$ and $b_5 \delta N^2 \delta K$, leaving us with
four independent transformations. Making use of eq.~\eqref{matter_var_diff-like-2}, the latter can be employed to set to zero four of the coefficients $a_I$ and $b_I$ for $I=6,7,8$, up to corrections that are at least third order in derivatives. This is summarized by Table \ref{table:3}.

In conclusion, the higher-derivative corrections to the leading order dynamics $\delta N^2$ and $\delta N^3$ start quadratic in $H/\Lambda$ and there are only 3 two-derivative corrections. 
This is a major simplification: out of 17 operators, 12 are redundant and one is left with only 5 of them.
It is again straightforward to consider higher-order operators. 
At each order one has 8 new operators, 2 $f$-like field redefinitions and 3 new $g$'s: only 3 couplings are not redundant.

\renewcommand{\arraystretch}{1.2}
\begin{table}
\begin{center}
\begin{adjustbox}{max width=\textwidth}
\begin{tabular}{ | c | l | c | c | c | l | c | }
\hline 
Coeff.	& ${\cal O}^{(2)}$			&	\#$\partial_\mu$	& $\to 0$	& $\to 0$		\\ [0.0cm] 
\hline 
\hline 
$a_0$	& $\tr$					&	2	& $f_{1,2}$	&  \\ [0.0cm] 
\hline 
$a_1$	& $\delta K_{\mu\nu}\delta K^{\mu\nu}$	& 2 & $f_{1,2}$ & 	\\ [0.0cm] 
\hline 
$a_2$	& $\tr\,\delta N$				&	2	& $f_{3,4}$ &  	\\ [0.0cm] 
\hline 
$a_3$	& $\AA_\mu \AA^\mu$		&	2	&		&	\\ [0.0cm] 
\hline 
$a_4$	& $H^2 \delta N^2$			&	0	&   & 	\\ [0.0cm] 
\hline 
$a_5$	& $H \delta N \delta K$		&	1	&   & $g_{1}$	\\ [0.0cm] 
\hline 
$a_6$	& $\delta K^2$				&	2	&   & $g_{5}$		\\ [0.0cm] 
\hline 
$a_7$	& $\VV^2$				&	2	&   & $g_{3}$		\\ [0.0cm] 
\hline 
$a_8$	& $\VV \delta K$			&	2	&    & $g_{3,5}$	\\ [0.0cm] 
\hline
\end{tabular}

\qquad

\begin{tabular}{ | c | l | c | c | c | l | c | }
\hline
Coeff.	& ${\cal O}^{(3)}$			&	\#$\partial_\mu$				& $\to 0$	 & $\to 0$		\\ [0.0cm] 
\hline
\hline 
	&  	& 	& 			&	\\ [0.0cm] 
\hline 
$b_1$	& $\delta N\delta K_{\mu\nu}\delta K^{\mu\nu}$  &	2		& $f_{3,4}$	& 				\\ [0.0cm] 
\hline
$b_2$	& $\tr\,\delta N^2$		&	2					& $f_{5,6}$	& 				\\ [0.0cm] 
\hline
$b_3$	& $\delta N\AA_\mu \AA^\mu$		&	2			& $f_{5}$		& 					\\ [0.0cm] 
\hline
$b_4$	& $H^2 \delta N^3$			&	0				&   &  				\\ [0.0cm] 
\hline
$b_5$	& $H \delta N^2 \delta K$		&	1				& $f_{5,6}$   &  $g_{2}$	\\ [0.0cm] 
\hline 
$b_6$	& $\delta N \delta K^2$			&	2			&  	& $g_{5,6}$				\\ [0.0cm] 
\hline
$b_7$	& $\delta N\VV^2$			&	2				&  	& $g_{3,4}$				\\ [0.0cm] 
\hline
$b_8$	& $\delta N \VV \delta K$		&	2				& $f_{5}$  &  $g_{3,4,5,6}$		\\ [0.0cm]
\hline
\end{tabular}
\end{adjustbox}
\end{center}
\caption{\footnotesize{Quadratic (left panel) and cubic (right panel) operators up to second order in derivatives. 
The fourth column 
shows which operator can be set to zero by the transformation \eqref{f1234}, which is exact in the derivative expansion, see Section~\ref{sec:operators}. Treating higher derivatives perturbatively, the fifth column 
shows which operator can be set to zero by the transformation \eqref{diff-like_field_redef}. The transformations $g_1$ and $g_2$ are used to set to zero the one-derivative operators $a_5$ and $b_5$, while $g_{3,4,5,6}$ can be used to set to zero four of the two-derivative operators $a_{6,7,8}$ and $b_{6,7,8}$. }}
\label{table:3}
\end{table}

\subsection{Higher-derivative operators for tensors}
\label{subs:tensor_higher}

We now consider operators with more than 2 derivatives on the metric. Instead of remaining general we focus on operators that modify the tensor dynamics, which cannot be changed at the two-derivative level.

\subsubsection{Operators with three derivatives}
\label{conti_gamma_gamma_gamma}

\noindent Possible three-derivative operators for tensors up to cubic order are 
\be \label{3-der_ops}
\tr_{\mu\nu}\delta K^{\mu\nu}\,, \quad \delta R^{\mu0\nu0}\delta K_{\mu\nu} \quad \text{and} \quad \delta K_{\mu\nu}\delta K^\mu_{\,\,\rho}\delta K^{\nu\rho}\;.
\ee
(Here we are assuming parity. For a discussion about parity violating operators one can see \cite{Creminelli:2014wna}.) However, using the relation \cite{Gleyzes:2013ooa} 
\be
\lambda(t) \tr_{\mu\nu}K^{\mu\nu} = \frac{\lambda(t)}{2}\tr K + \frac{\dot\lambda(t)}{2N}\tr + \text{boundary terms}\;,
\ee
one can get rid of the first operator. Moreover, using the Gauss-Codazzi relation, one can show that
\be
N^2 \, K_{\alpha\gamma}R^{\alpha 0\gamma 0} = - K^{\alpha\gamma}K_\gamma^{\,\,\rho}K_{\rho\alpha} + K_\alpha^{\,\,\gamma} D_\gamma A^\alpha + K_\alpha^{\,\,\gamma} A_\gamma A^\alpha - K_\alpha^{\,\,\gamma} n^\delta\nabla_\delta K_\gamma^{\,\,\alpha} \;.
\ee
The second and third operators contain scalar perturbations, while the last one can be integrated by parts.
In this way one can also dispose of the second operator in eq.~\eqref{3-der_ops}. 
One can then wonder whether it is possible to set also the third operator to zero with a suitable field redefinition: 
as we are now going to show, this is not possible.

To see this, one has to find all the possible field redefinitions that carry one derivative on the metric. 
Since the only scalars that satisfy this requirement are $K$ and $V$, 
and the only symmetric tensors that we can add to $g_{\mu\nu}$ are $K_{\mu\nu}$ and $n_{(\mu} A_{\nu)}$, 
we see that eq.~\eqref{general_transf} reduces to 
\begin{equation}
\label{eq:general_transf_red}
g_{\mu \nu} \to C(t,N,K,V) g_{\mu \nu} + D(t,N,K,V) n_\mu n_\nu + E(t,N) \delta K_{\mu\nu} + F(t,N)n_{(\mu} A_{\nu)}\;,
\end{equation}
where we have considered $\delta K_{\mu\nu}$ instead of $K_{\mu\nu}$ on the r.h.s., without loss of generality. 
Since $A_\mu$ does not contain tensor modes, the term $\propto n_{(\mu}A_{\nu)}$ cannot affect the cubic action for three gravitons. 
Therefore, the only way to possibly induce the operator $\delta K_{\mu\nu}\delta K^\mu_{\,\,\rho}\delta K^{\nu\rho}$ is a transformation of the form 
\be
\label{eq:cK_field_redefinition-A}
g_{\mu\nu} \to g_{\mu\nu} + c_K \delta K_{\mu\nu}\;.
\ee 
It is now straightforward to see that we cannot generate $\delta K_{\mu\nu}\delta K^\mu_{\,\,\rho}\delta K^{\nu\rho}$ through eq.~\eqref{eq:cK_field_redefinition-A}. 
When written in terms of $\gamma_{ij}$, the transformation becomes 
\be
\label{eq:cK_field_redefinition-B}
\gamma_{ij} \to \gamma_{ij} + c_K\dot\gamma_{ij}\;,
\ee
i.e. a linear shift at all orders in perturbations. 
Therefore, the only effect it has is to change $S^{(2)}_{\gamma}$ and $S^{(3)}_{\gamma}$ separately: 
since in the Einstein-Hilbert action $S^{(3)}_{\gamma}$ comes only from $\tr$ \cite{Maldacena:2002vr, Maldacena:2011nz}, 
at leading order in $c_K$ we will just have generated terms with two spatial derivatives and one time derivative, 
and no terms of the form $\dot{\gamma}_{ij}^3$.

\subsubsection{Operators with four derivatives}
\label{four_derivs_section}

\noindent 
There are no parity-conserving corrections to the tensor power spectrum with three derivatives.
The first correction is at fourth order in derivatives, \cite{Cannone:2014uqa}.
Up to integration by parts, there are four operators with four derivatives that modify the tensor power spectrum:
\be
\label{4_der_ops}
\tr_{\mu\nu}^2\,, \quad (\nabla^0\delta K_{\mu\nu})^2\,, \quad \tr_{\mu\nu}\nabla^0\delta K^{\mu\nu} \quad \text{and} \quad (D_\rho\delta K_{\mu\nu})^2\;.
\ee
The corresponding modifications in the quadratic action for tensors are of the form $(\d^2\gamma_{ij})^2$, for the first, $\ddot \gamma_{ij}^2$ for the second and $(\d_k\dot\gamma_{ij})^2$ for the last two.
One has the freedom to perform field redefinitions, but there are not enough of them to get rid of all the three operators. 
Indeed, there are only two possible field redefinitions at second order in derivatives that affect tensor modes:
\be \label{4_der_redef}
g_{\mu\nu} \to g_{\mu\nu} + c_R \tr_{\mu\nu} + c_0 \nabla^0 \delta K_{\mu\nu}\;.
\ee
They correspond to
$\gamma_{ij} \to c_R \, \d^2 \gamma_{ij} + c_0 \, \ddot \gamma_{ij}$.
We conclude that we cannot eliminate all the corrections to the tensor power spectrum at this order.

The modification of the power spectrum is only possible because of the preferred foliation provided by the inflaton. 
In the absence of a preferred foliation one is forced to write only operators that are fully diffeomorphism-invariant. 
Since the Gauss-Bonnet term, $R^2-4 R^{\mu\nu} R_{\mu\nu}+R^{\mu\nu\rho\sigma}R_{\mu\nu\rho\sigma}$, is a total derivative in four dimensions, the only allowed operator is $\delta R_{\mu\nu}^2$. 
However, one can dispose of it by using the field redefinition $g_{\mu\nu} \to g_{\mu\nu}+ \delta R_{\mu\nu}$.

Let us now compute the correction to the tensor power spectrum due to the new couplings. 
To simplify things, we use the free parameters $c_R$ and $c_0$ in eq.~\eqref{4_der_redef} to set to zero the coupling in front of $(\nabla^0\delta K_{\mu\nu})^2$ and in front of the quadratic part of the combination $c_1 \tr_{\mu\nu}\nabla^0\delta K^{\mu\nu}+ c_2 (D_\rho\delta K_{\mu\nu})^2$. 
Therefore, we just need to expand $\tr_{\mu\nu}^2$ at quadratic order in $\gamma$. We get 
\be
\label{S2_from_3Rij2}
S_4^{(2)} = \frac{1}{4} \frac{M_{\rm Pl}^2}{\Lambda^2} \int\dif\eta\,\dif^3x\,(\d^2\gamma_{ij})^2\;.
\ee
It is straightforward to study the effect of this term in the usual in-in formalism. 
The interaction Hamiltonian $\mathcal H_\text{int}$ in Fourier space is
\be
\mathcal H_\text{int} = {- \frac{1}{4} \frac{M_{\rm Pl}^2}{\Lambda^2}} \int\frac{\dif^3k}{(2\pi)^3}\,k^4\sum_{s_1,s_2} \gamma^{s_1}_{\vec{k}_1} \gamma^{s_2}_{\vec{k}_2}\, \ep_{ij}^{s_1}({\vec{k}_1})\ep_{ij}^{s_2}({\vec{k}_2})\;.
\ee
Then, the correction to the power spectrum is given by
\be 
\delta\braket{\gamma^{s}_{\vec{k}}\,\gamma^{s'}_{-\vec{k}}}' = \frac{5}{4}\frac{H^2}{\Lambda^2}\frac{H^2}{2 M_{\rm Pl}^2 k^3}\delta_{s s'}\;,
\ee
where the $'$ means that we omitted the factor $(2\pi)^3$ and the Dirac delta.
In general, one expects the above correction to be small, being suppressed by a factor $H^2/\Lambda^2$. 
However it could become sizable if the suppression scale $\Lambda$ is not too large. 

When the power spectrum is modified, we also expect sizable non-Gaussianities. 
More precisely, we expect an enhancement if we consider three-point functions involving scalars. 
In fact, the operators \eqref{4_der_ops} are constructed from the foliation, i.e.~they entail a direct coupling with the inflaton.
For instance, we expect that their contribution to the cubic action $\gamma\gamma\zeta$ will not be suppressed by slow-roll parameters. On the other hand, a slow-roll suppression is present at the 2-derivative level where, as we discussed in Section \ref{conti_gamma_gamma_zeta}, the only freedom comes by the modification of the scalar constraint equations.

We can estimate these non-Gaussianities in the following way. 
At cubic order in perturbations, 
given that we have used $c_R$ and $c_0$ to put the quadratic action in the form of eq.~\eqref{S2_from_3Rij2}, 
there will be $\tr_{\mu\nu}^2$ and the cubic part of $c_1 \tr_{\mu\nu}\nabla^0\delta K^{\mu\nu}+ c_2 (D_\rho\delta K_{\mu\nu})^2$ that will contribute. 
For our estimation, however, it is enough to consider $\tr_{\mu\nu}^2$. 
In Appendix~\ref{zeta_gamma_gamma_bispectrum} we show that, 
as expected, for two gravitons and a scalar we have 
$S_{4}^{(3)}\sim\frac{H^2}{\Lambda^2}\times\ep^0$. 
Since $S_0^{(3)}\sim\ep$ \cite{Maldacena:2002vr}, 
we see that the bispectrum coming from these $4$-derivative operators could dominate the standard slow-roll result for $\frac{H^2}{\Lambda^2}\gtrsim\ep$.

\section{Field redefinitions in the decoupling limit}
\label{sec:decoup_limit}

\noindent In many cases inflationary correlation functions can be calculated, at leading order in slow-roll, in the so-called decoupling limit. This means concentrating on the Goldstone field $\pi$, which is introduced in the action when we depart from the unitary gauge, and neglecting the fluctuations of the metric. 
In this limit one can consider field redefinitions of $\pi$ that decay at late times and thus do not change the asymptotic correlation functions. A natural question is whether these field redefinitions of $\pi$ are simply the decoupling limit of the ones we discussed before or they are different in nature. We want to argue that, in general, these two kinds of field redefinitions are different and cannot be simply related. 

First of all, notice that the interactions of $\pi$ are constrained by the non-linear realization of Lorentz invariance. Indeed if we neglect metric perturbations and we go to short scales, we have a theory with spontaneously broken Lorentz symmetry: the combination $t + \pi(t,\vec x)$ transforms as a scalar under Lorentz and this defines the non-linear transformation of $\pi$. Starting from a generic unitary gauge action and reintroducing $\pi$ with the usual Stueckelberg procedure, one ends up in the decoupling limit with an action of $\pi$ with this well-defined non-linear realization of the Lorentz symmetry. In particular, this will remain true even when one performs a metric field redefinition in unitary gauge and considers two equivalent actions in the sense discussed above: in the decoupling limit of either theories one has the same non-linear realization of the Lorentz symmetry. This, however, does not happen when one considers a general field redefinition of $\pi$ in the decoupling limit:
\be
\tilde \pi = \pi + f(\pi, \dot \pi, \partial_i \pi, \ldots) \;.
\ee
Given the transformation rules of $\pi$, one sees that $\tilde \pi$ will transform in a different way for a generic $f$. This is enough to show that the action for $\tilde \pi$ cannot generically be obtained as the decoupling limit of an action in unitary gauge via Stueckelberg. 

Let us focus on a concrete example where a $\pi$ field redefinition is useful. In single-field inflation the leading operators giving a potentially large 3-point function for $\zeta$ are $\dot\pi^3$ and $\dot\pi (\partial \pi)^2$. At subleading order in derivatives one should look at cubic operators with four derivatives. It is straightforward to realize that (up to integration by parts) there are only two 4-derivative operators \cite{Creminelli:2010qf, Behbahani:2014upa}:
$\partial^2\pi (\partial \pi)^2$ and $\partial^2\pi\, \dot\pi^2$. They arise from the unitary gauge operators $\delta N \delta K$ and $\delta N^2 \delta K$. The action in the decoupling limit is given by
\be
\begin{split}
\label{dlaction}
&S_\pi = \int \dif^4 x\, a^3\,(-\mpl^2 \dot H) \bigg[ (1+\alpha_1)\(\dot\pi^2-c_\mathrm{s}^2 \frac{(\partial\pi)^2}{a^2}\) 
+ \left( {\alpha_2}-\alpha_1 \right) \dot\pi\frac{(\partial\pi)^2}{a^2} \\
&\hphantom{S_\pi = \int \dif^4 x\, a^3\,(-\mpl^2 \dot H) \bigg[ } 
- 2 ( \alpha_1 + \alpha_3) \dot\pi^3 + 2 \frac{ \alpha_2 - \alpha_4}{H} ~\dot\pi^2 \frac{\partial^2\pi}{a^2} + \frac{\alpha_2}{H} ~\frac{(\partial\pi)^2\partial^2\pi}{a^4} \bigg]\;,
\end{split}
\ee
with $c_\mathrm{s}^2 \equiv (1 + \alpha_2)/(1 + \alpha_1)$. We use here the notation of \cite{Pirtskhalava:2015zwa}; the $\alpha$'s are related to our $a$'s as $\alpha_1 = - a_4/\epsilon\,,\,\,\alpha_2 = - a_5/2\epsilon\,,\,\,\alpha_3 = - b_4/2\epsilon\,,\,\,\alpha_4 = b_5/2\epsilon\,$. 
Naively, the last two operators in the equation above give 3-point functions whose shape is different from the standard operators with three derivatives. 
However, it is straightforward to check that the field redefinitions $\pi \to \pi + c_1 (\partial \pi)^2$ and $\pi \to \pi + c_2 \dot\pi^2$ can be used to remove both these operators, while changing the coefficient of the 3-derivative operators $\dot\pi^3$ and $\dot\pi (\partial \pi)^2$. This shows that, in the decoupling limit, operators with one extra derivative do not give rise to new shapes.\footnote{The same argument can be run at any order in $\pi$ to argue that there are no genuine new terms with one extra derivative at any order in $\pi$.} The removal of the 4-derivative term does not mean the theory is equivalent to one with only $\delta N^2$ and $\delta N^3$. This can be seen noting that the operator $\dot\pi (\partial \pi)^2$, after the field redefinition, has a coefficient $\alpha_2-\alpha_1 + 2\alpha_2 (1+\alpha_1)/(1+\alpha_2)$, which is not related to $c_\mathrm{s}$ in the standard way as dictated by the non-linear realization of Lorentz invariance. 

In Section \ref{diff-like_trans} we showed that the operators $\delta N \delta K$ and $\delta N^2 \delta K$ can be removed by a unitary-gauge field redefinition, provided their coefficients are small so that one can neglect quadratic corrections. 
The corresponding statement in the decoupling limit should be that there is a field redefinition of $\pi$ which preserves the usual Lorentz transformation of $\pi$ and gets rid of the 4-derivative terms at linear order in their coefficient. 
Since the combination $\psi \equiv t + \pi$ transforms as a Lorentz scalar, also $(\partial_\mu\psi)^2 + 1 = - 2 \dot\pi - \dot\pi^2 +(\partial\pi)^2$ transforms as a scalar. This means that the field redefinitions
\be
\tilde \pi = \pi + c_1 (2 \dot\pi + \dot\pi^2 -(\partial\pi)^2)\;, \qquad \tilde \pi = \pi + c_2 (2 \dot\pi + \dot\pi^2 -(\partial\pi)^2)^2 = \pi + 4 c_2 \dot\pi^2 + \ldots 
\ee
preserve the Lorentz transformations of $\pi$. Notice that there is now a linear piece in the first transformation: this means one has to restrict the transformation to linear order in $c_1$ to avoid the proliferation of higher derivatives and that the quadratic action, and in particular the speed of sound, will be modified. 
It is straightforward to check that using these field redefinition one can eliminate the 4-derivative terms in eq.~\eqref{dlaction} {\em at linear order in $\alpha_2$ and $\alpha_4$} and that the resulting theory has the usual relation between the speed of sound and the coefficient of the operator $\dot\pi (\partial \pi)^2$. 
This is the decoupling limit of a unitary gauge action in which $\delta N \delta K$ and $\delta N^2 \delta K$ are removed. 

In conclusion, the $\pi$ field redefinitions in the decoupling limit is an extra freedom that one is not able to trace in the unitary gauge theory. 
This should not be surprising after all: in the case of a spontaneously broken non-abelian gauge theory, one has freedom to parametrize the coset of the Goldstones in various way. 
This freedom has no obvious analogy in unitary gauge where the Goldstones are eaten by the massive gauge fields.

\section{Conclusions}
\label{conclusions}

\noindent In this paper we have explored the effect of generalized disformal transformations in the Effective Field Theory of Inflation. These transformations do not change the predictions for the late-time observables and can thus be used to simplify the action. They can be organized in an expansion in derivatives and perturbations. These are the main results we obtained.
\begin{itemize}[leftmargin=*]
\item If one considers (unitary gauge) operators with up to two derivatives and up to $n$-th order in perturbations ($n \geq 2$), one has $8(n-1)+1$ independent operators (taking into account integration by parts). 
$2n$ of these can be set to zero by conformal and disformal transformations, which carry powers of $\delta N$ up to $\delta N^{n-1}$.
\item Using these transformations, it is easy to show that the predictions for the tensor power spectrum and the correlator $\braket{\ggg}$ cannot be modified at leading order in derivatives \cite{Creminelli:2014wna}. 
Also all the couplings contributing to $S^{(3)}_{\ggz}$ beyond the Einstein-Hilbert action can be removed. 
Even so, $\braket{\ggz}$ will still be affected by the possible changes in the scalar sector through the constraint equations, therefore we cannot conclude that $\braket{\ggz}$ is fully fixed.
\item Among the additional transformations that contain derivatives, some do not affect the Einstein-Hilbert action but only the inflaton part. 
These can be used to reduce the number of higher-derivative corrections. 
For instance, if one starts from a theory where the dominant terms in the inflaton action are those with zero derivatives, one has six additional transformations (up to cubic order in perturbations) that can be used to further simplify the action. One is left with only three higher-derivative corrections up to 2-derivative order.
\item At 3-derivative order, there are no corrections to the tensor power spectrum and only one independent operator contributing to $S^{(3)}_{\ggg}$ after integration by parts.
\item At 4-derivative order, there is only one independent operator that affects the tensor power spectrum. This is due to the coupling with the inflaton and as such can give a large bispectrum $\expect{\gamma\gamma\zeta}$. 
\item In the decoupling limit, one can perform field redefinitions of the Goldstone $\pi$ to simplify the action. In general this kind of transformations does not preserve the way $\pi$ transforms under Lorentz and cannot be seen as the decoupling limit of the unitary gauge transformations discussed above.
\end{itemize}

It would be interesting to understand how to phenomenologically identify the few higher-derivative corrections we are left with after the field redefinitions.
Also the potentially large bispectrum $\expect{\gamma\gamma\zeta}$ due to four-derivative operators deserves further studies.

\section*{Acknowledgements}

\noindent It is a pleasure to thank Daniel Baumann, Andrei Khmelnitsky, Michele Mancarella, Enrico Pajer, Guilherme 
Pimentel, Cora Uhlemann and especially Mehrdad Mirbabayi for useful discussions. 
G.C.~acknowledges support by the grant Theoretical Astroparticle Physics number 2012CPPYP7, 
under the program PRIN 2012 funded by MIUR and by TASP (iniziativa specifica INFN). 
F.V.~acknowledges partial financial support from ``Programme National de Cosmologie and Galaxies'' (PNCG) of CNRS/INSU, France.

\appendix

\section{Transformation of the operators \texorpdfstring{$\mathcal{O}^{(2)}_I$ and $\mathcal{O}^{(3)}_I$}{O\^{}(2)\_I and O\^{}(3)\_I}}
\label{app:transformations}

\noindent In this appendix we compute how the couplings $a_I$ and $b_I$ change under the transformations $f_i$.
The transformation of the metric that we are considering is of the form
\begin{equation}
\label{eq:app_A-transformation_active}
g_{\mu\nu}\to\widetilde{g}_{\mu\nu} = C(t,N) g_{\mu\nu} + D(t,N) n_\mu n_\nu\;.
\end{equation}
This amounts to a rescaling of the three-metric and the normal one-form: 
more precisely, we have $\widetilde{h}_{\mu\nu} = C(t,N)h_{\mu\nu}$, $\widetilde{n}_\mu = \sqrt{C(t,N)-D(t,N)}\,n_\mu$.
Recalling that $n_\mu = -N\dif t_\mu$, 
this amounts to a rescaling of the lapse function $\widetilde{N} = \sqrt{C(t,N)-D(t,N)}\,N$. 
In order to see how the coefficients $a_I$ and $b_I$ transform, 
we need to invert these transformations, that is
\begin{subequations}
\label{eq:app_A-transformation_passive-A}
\begin{align}
h_{\mu\nu} &= \frac{\widetilde{h}_{\mu\nu}}{C(t,N)}\;, \label{eq:app_A-transformation_passive-A-1} \\
n^\mu &= \sqrt{C(t,N)-D(t,N)}\,\widetilde{n}^\mu\;. \label{eq:app_A-transformation_passive-A-2}
\end{align}
\end{subequations}
We also need to do a time rescaling to bring the background value of $\widetilde{g}_{00}$ to one for the unperturbed transformations $f_1$ and $f_2$. 
That is, we want $N = 1$ for $\widetilde{N} = 1$. 
To find the expression for the rescaling, we use the following facts:
\begin{itemize}[leftmargin=*]
\item Standard results of the ADM decomposition (see, e.g., \cite{Wald:1984rg}) tell us that
\begin{equation}
\label{eq:app_A-transformation_passive-B}
N = (\dif t_\mu n^\mu)^{-1} = \frac{(\dif t_\mu \widetilde{n}^\mu)^{-1}}{\sqrt{C(t,N)-D(t,N)}}\;.
\end{equation}

\item The normal $\widetilde{n}^\mu$ does not change under a time rescaling. 
\end{itemize}
From this, we see that a redefinition $\dif t_\mu = \left[\widebar{C}(t(\tilde{t}\,)) - \widebar{D}(t(\tilde{t}\,))\right]^{-1/2}\,\dif\tilde{t}_\mu$, where a bar denotes a background quantity, does the job. 
We then \mbox{arrive at} 
\begin{equation}
\label{eq:app_A-transformation_passive-C}
N = \frac{\sqrt{\widebar{C}(t(\tilde{t}\,)) - \widebar{D}(t(\tilde{t}\,))}}{\sqrt{C(t(\tilde{t}\,),N(\widetilde{N})) - D(t(\tilde{t}\,),N(\widetilde{N}))}}\widetilde{N}\;.
\end{equation}
This relation between $N$ and $\widetilde{N}$ can be expanded around $1$ in perturbations, 
and can be used to solve perturbatively for $C$ and $D$ in terms of $\widetilde{\delta N}$. 
With some abuse of notation, 
in the following we will use $C$ and $D$ to mean the conformal and disformal factors solved in terms of $\widetilde{\delta N}$. 
The final two things we need are how the volume element $\dif^4x\,\sqrt{-g}$ and the Hubble factor transform. 
The transformation of the volume element is straightforward to compute, 
since it 
is not affected by the time rescaling. 
The relation between $H$ and $\widetilde{H}$ follows from the change in the scale factor due to $\widebar{C}$, 
and the time rescaling. 
It is given by (we will also suppress the time arguments of $C$ and $D$, in the following)
\begin{equation}
\label{eq:app_A-hubble_change}
H = \sqrt{\widebar{C}(t(\tilde{t}\,)) - \widebar{D}(t(\tilde{t}\,))}\bigg[\widetilde{H} - \frac{1}{2}\frac{\dif\log\widebar{C}(t(\tilde{t}\,))}{\dif\tilde{t}}\bigg]\;.
\end{equation}
These results allow us to compute how a generic operator 
transforms under \eq{app_A-transformation_passive-A}:
\begin{itemize}[leftmargin=*]
\item Let us start from the transformation of $\delta K_{\mu\nu}$. We have that 
\begin{equation}
\label{eq:Kmunu-A}
\delta K_{\mu\nu} = K_{\mu\nu} - Hh_{\mu\nu} = K_{\mu\nu} - \frac{H(\widetilde{H})}{C}\tilde{h}_{\mu\nu}\;,
\end{equation}
so we just need to see how $K_{\mu\nu}$ transforms. Recall that 
\begin{equation}
\label{eq:Kmunu-B}
K_{\mu\nu} = \frac{1}{2}\mal{L}_{\vec{n}}h_{\mu\nu}\;,
\end{equation}
so it is straightforward to plug in this formula the relation of $h_{\mu\nu}$ and $n^\mu$ to $\widetilde{h}_{\mu\nu}$ and $\widetilde{n}^\mu$ to arrive at
\begin{equation}
\label{eq:Kmunu-C}
\begin{split}
&K_{\mu\nu} = \frac{1}{2}\mathcal{L}_{\sqrt{C-D}\,\widetilde{\vec{n}}}\left(C^{-1}\widetilde{h}_{\mu\nu}\right) = 
\frac{\sqrt{C-D}}{C}\left[\widetilde{K}_{\mu\nu} + \left(\mathcal{L}_{\widetilde{\vec{n}}}\log C^{-1/2}\right)\widetilde{h}_{\mu\nu}\right]\;.
\end{split}
\end{equation}
Expanding $\widetilde{K}_{\mu\nu}$ as $\widetilde{\delta K}_{\mu\nu} + \widetilde{H}\widetilde{h}_{\mu\nu}$, 
and plugging it in \eq{Kmunu-A}, 
we arrive at the transformation of $\delta K_{\mu\nu}$. 
Notice that the terms $\propto\mathcal{L}_{\widetilde{\vec{n}}}\log C$ will give rise to terms $ \widetilde{\delta N}^n\times\widetilde{V}$, which must be integrated by parts to stay in Table \ref{table:quadratic} and Table \ref{table:cubic}.

\item The transformation of $\tr\,\delta N^n$ includes a straightforward conformal transformation. 
We point out that terms $ \widetilde{D}_\mu \widetilde{D}^\mu\widetilde{\delta N}$ will be generated, 
which must also be integrated by parts to yield $ \widetilde{A}_\mu\widetilde{A}^\mu$.

\item Finally, we list the transformation properties of $V$ and $A_\mu = D_\mu\log N$:
\begin{subequations}
\label{eq:V_and_A}
\begin{align}
&V = \sqrt{C-D}\left[\widetilde{V} + \widetilde{N}\,\widetilde{n}^\mu\widetilde{\nabla}_\mu{\frac{\sqrt{\widebar{C} - \widebar{D}}}{\sqrt{C - D}}}\right]\;, \label{eq:V_and_A-1} \\
&A_\mu = \widetilde{A}_\mu - \widetilde{D}_\mu\log\sqrt{C-D}\;. \label{eq:V_and_A-2}
\end{align}
\end{subequations}
\end{itemize}

These relations can be used to derive the transformations of the coefficients $a_I$ and $b_I$. 
We start from the action eq.~\eqref{general_action} where $a_0$ and $a_1$ have been set to zero by the use of $f_1$ and $f_2$, as discussed in Section~\ref{sec:operators} \cite{Creminelli:2014wna}.
For simplicity, we assume that the coefficients $a_I$ and $b_I$ are time independent and we first consider the effect of $f_3$ and $f_4$ only on the operators that are not affected by $f_5$ and $f_6$. 
Moreover, for convenience we define
\be
\Coe \equiv \frac{1}{1+\frac{f_3}{2} - \frac{f_4}{2}} \;.
\ee
Here we assume that $f_3$ and $f_4$ are time independent but not necessarily small. In particular, the transformations are non-linear in these two parameters.
With these assumptions, 
the action $\widetilde{S}$ will be of the same form of eq.~\eqref{general_action}, 
with coefficients $\widetilde{a}_I$, $\tilde{b}_I$ for the actions $\widetilde{\mathcal{L}}^{(2)}$ and $\widetilde{\mathcal{L}}^{(3)}$ and a $\widetilde{S}_0$ given by the standard Einstein-Hilbert plus the minimal inflaton action with coefficient $\widetilde{M}^2_\mathrm{Pl}/2 = M^2_\mathrm{Pl}/2$. 
The operator coefficients, obtained by writing eq.~\eqref{general_action} in terms of the metric on the r.h.s. of eq.~\eqref{f1234}, read
\begin{subequations}
\begin{align}
&\widetilde a_2 = \Coe \left[ a_2 + \frac12 \left(- f_3 + \frac{f_4}{2} \right) \right] \;, \\
&\widetilde a_3 = {\Coe^2} \left[ a_3 - 2 f_3 a_2 - f_3 \left(1 - \frac{f_3}{4} \right) \right]\;, \\ 
&\widetilde a_4 = {\Coe^2} \bigg\{ \left[ a_4 + \frac34 \left(f_3-f_4\right) \left(\left(4 f_3-f_4\right) \left(3 a_6-1\right)-3 a_8\right) + \frac34 \left(5 f_3-2 f_4\right) a_5\right] \nonumber \\
&\hphantom{\widetilde a_4 = {\Coe^2} \bigg\{ } + \frac{\dot{H}}{H^2} \left[ {-1} + \frac34 \left(f_3-f_4\right) \left( f_3 \left(3 a_6 - 1\right)-a_8\right)+ \frac34f_3 a_5 \right] \bigg\} + \frac{\dot H}{H^2} \;, \\ 
&\widetilde a_5 = {\Coe} \left[ a_5 + ( f_3 -f_4) ( 3 a_6 - 1) \right] \;, \\ 
&\widetilde a_7 = {\Coe^2} \left[ a_7 + \frac{3 f_3^2}{4 } ( 3 a_6 - 1) - \frac{3f_3}{2} a_8 \right] \;, \\
&\widetilde a_8 = {\Coe} \left[ a_8 - f_3 ( 3 a_6 - 1) \right] \;, \\
&\tilde b_1 = {\Coe} \left[ b_1 - \frac12 \left( f_3 + \frac{f_4}{2} \right) \right] \;, \\
&\tilde b_6 = {\Coe} \left[ b_6 - \frac12 \left( f_3 + \frac{f_4}{2} \right) ( 2 a_6 - 1 ) \right] \;,
\end{align}
\end{subequations}
while $a_6$ does not transform, i.e.~$\widetilde a_6 = a_6$.
As explained in the main text, one can choose $f_3$ and $f_4$ to set $\widetilde a_2$ and $\tilde b_1$ to zero, i.e.,
\be
\widetilde a_2=0 \;, \qquad \tilde b_1 = 0 \qquad \Longleftrightarrow \qquad f_3 = a_2 + b_1 \;, \qquad f_4=-2(a_2 - b_1) \;,
\ee
which changes the other coefficients according to the transformations above.
We can then explicitly compute the effect of $f_5$ and $f_6$ on the other operator coefficients, assuming that they are time independent and that $f_3=f_4=0$. In this case we have 
\begin{subequations}
\begin{align}
&\tilde b_2 = b_2 - \frac12 (f_5-f_6) a_2 - \frac14 \left( 2 f_5 - f_6 \right) \;, \\
&\tilde b_3 = b_3 - 4 f_5 a_2 -2 (f_5 -f_6) a_3 - 2 f_5 \;, \\ 
&\tilde b_4 = b_4 - (f_5-f_6) \left( a_4 + \frac32 a_8 \right) + \frac32 (3 f_5 -f_6) a_5 + \frac{\dot{H}}{H^2} \left[ (f_5-f_6) \left( 1 -\frac12 a_8 \right) + f_5 a_5 \right] \;, \\ 
&\tilde b_5 = b_5 - \frac12 (f_5 -f_6) (a_5 - 6 a_6 + 2 ) \;, \\
&\tilde b_7 = b_7 - 2 (f_5 - f_6) a_7 -3 f_5 a_8 \;, \\
&\tilde b_8 = b_8 - 6 f_5 a_6 - (f_5 -f_6) a_8 +2 f_5 \;.
\end{align}
\end{subequations}
This shows that $f_5$ can be used to set $b_2$, $b_3$, $b_5$ or $b_8$ to zero, while $f_6$ can be used only to set to zero $b_2$ or $b_5$.

\section{\texorpdfstring{$\expect{\gamma\gamma\zeta}$}{<\textbackslash gamma\textbackslash gamma\textbackslash zeta>} from \texorpdfstring{$\tr_{\mu\nu}^2$}{3R\_\{\textbackslash mu \textbackslash nu\}\^{}2}}
\label{zeta_gamma_gamma_bispectrum}

\noindent In this section we compute the $\gamma\gamma\zeta$ cubic action, and the corresponding bispectrum, associated to the change of the tensor power spectrum discussed in Section~\ref{four_derivs_section}. 
As explained in the main text, we focus on $\tr_{\mu\nu}^2$. After integration by parts, we find that the cubic action is equal to 
\begin{equation}
\begin{split}
&S_4^{(3)} \supseteq \frac{\mpl^2}{\Lambda^2} \int\dif\eta\,\dif^3 x\, \bigg[ \d_i\d_j \zeta\, \d^2 \gamma_{ik}\, \gamma_{kj} + \frac{1}{2}\d^2 \zeta\,\d_i \gamma_{kl}\, \d_i\gamma_{kl} + \frac{1}{2} \d_i\d_j \zeta\, \d_i \gamma_{kl}\,\d_j \gamma_{kl} \\
&\hphantom{S_4^{(3)} \supseteq \frac{\mpl^2}{\Lambda^2} \int\dif\eta\,\dif^3 x\, \bigg[ } + \d_i\d_j \zeta\, \d_i\d_l\gamma_{kj}\, \gamma_{kl} - \frac{1}{4} \zeta\, \d^2\gamma_{ij}\, \d^2 \gamma_{ij} + \frac{1}{2} \d_k \zeta\, \d_k\gamma_{ij}\, \d^2 \gamma_{ij} \bigg]\;,
\end{split}
\end{equation}
i.e. it is not slow-roll suppressed, as expected. Using the in-in formalism one can easily compute the associated three-point function. It is equal to
\begin{equation}
\begin{split}
\langle \zeta_{\vec{k}_1}\gamma^{s_2}_{\vec{k}_2}\gamma^{s_3}_{\vec{k}_3} \rangle ' = & \ \frac{H^2}{\Lambda^2} \frac{1}{\ep} \left(\frac{H}{M_{\rm Pl}}\right)^4 \left[ {\cal I} (\k_1,\k_2,\k_3) + {\cal I} (\k_1,\k_3,\k_2) \right] \\
& 
\, \times \frac{1}{(k_1\,k_2\,k_3)^3} \ \frac{1}{K^4}\bigg( \sum_i k_i^3 + 4 \sum_{i\neq j}k_i k_j^2 + 12\, k_1 k_2 k_3 \bigg)\;,
\end{split}
\end{equation}
where $K = k_1+k_2+k_3$. The function $\mal{I}$ is, instead, given by 
\begin{equation}
\begin{split}
&\mathcal I(\k_1,\k_2,\k_3) = \bigg\{ k_2^2\, (\k_1\cdot\ep^2\cdot\ep^3\cdot\k_1) + \k_1\cdot\k_2\, (\k_1\cdot\ep^2\cdot\ep^3\cdot\k_2) \\
&\hphantom{\mathcal{I}(\k_1,\k_2,\k_3) = \bigg\{ } + \frac{1}{2} \left( k_1^2\, (\k_2\cdot\k_3) + (\k_1\cdot\k_2)\,(\k_1\cdot\k_3) - \frac{1}{2} k_2^2\,k_3^2 + k_3^2\, (\k_1\cdot\k_2) \right) [\ep^2\cdot\ep^3] \bigg\}\;,
\end{split}
\end{equation}
where $\ep^i = \ep_{ij}^{s_i}(\k_i)$ and $[\,\cdot\,]$ denotes the trace.


\clearpage


\bibliographystyle{utphys}
\bibliography{EFT_I_refs-gio}

\end{document}